\documentclass[prd,amsmath,amssymb,preprintnumbers,nofootinbib, preprint
,longbibliography]{revtex4-1}
\pdfoutput=1
\usepackage[utf8]{inputenc} 
\usepackage{graphicx}
\usepackage{url}
\usepackage[bookmarks, pagebackref=false]{hyperref}
\usepackage[usenames,dvipsnames]{xcolor}
\usepackage{cleveref}

\definecolor{orange}{cmyk}{0,0.5,1,0}
\definecolor{graa}{rgb}{0.8,0.8,0.8}
\definecolor{blaa}{rgb}{0.2,0.2,0.6}

\definecolor{colA}{HTML}{c19277}
\definecolor{colB}{HTML}{e1bc91}
\definecolor{colD}{HTML}{62959c}

\hypersetup{
  colorlinks, 
  bookmarksopen, 
  bookmarksnumbered,
  citecolor=colA, 		
  linkcolor=colA,	
  urlcolor=colD,			
}
\usepackage{amsthm}
\usepackage{bm}
\usepackage{bbm}
\usepackage{pxfonts}

\usepackage{amsmath,amssymb,amsfonts}
\usepackage{color}
\usepackage{hyperref}
\usepackage[Symbolsmallscale]{upgreek}
\usepackage{amsmath}
\usepackage{amsfonts}
\usepackage{amssymb,dsfont}
\usepackage{graphicx}
\usepackage{amssymb}
\usepackage{bm}

\usepackage{placeins}
\usepackage{xspace}
\usepackage{cancel} 
\usepackage{slashed}
\usepackage{natbib}

\usepackage{tikz}
\usetikzlibrary{intersections}
\usetikzlibrary{arrows,decorations.pathmorphing,backgrounds,positioning,fit,petri}
\usetikzlibrary{decorations.markings}
\usetikzlibrary{calc}
\usetikzlibrary{intersections,through,hobby}

\usepackage{diagbox}
\newcommand{\MS}{\ensuremath{\overline{\mathrm{MS}}}}
\newcommand{\ep}{\ensuremath{\varepsilon}}
\newcommand{\KRS}{\ensuremath{\mathcal{KR}^*}}
\newcommand{\KRP}{\ensuremath{\mathcal{KR}^\prime}}
\newcommand{\boxL}[1]{\tikz[baseline]{\node[draw=red!25, anchor=base, line width=1pt]{$#1$};}}
\newcommand{\boxN}[1]{\tikz[baseline]{\node[draw=ForestGreen!25, anchor=base, line width=1pt]{$#1$};}}
\newcommand{\boxJ}[1]{\tikz[baseline]{\node[draw=MidnightBlue!25, anchor=base, line width=1pt]{$#1$};}}
\newcommand{\boxLJ}[1]{\tikz[baseline]{\node[draw=MidnightBlue!25, anchor=base, line width=1pt]{\tikz[baseline]{\node[draw=red!25, anchor=base, line width=1pt]{$#1$};}};}}
\newcommand{\boxNJ}[1]{\tikz[baseline]{\node[draw=MidnightBlue!25, anchor=base, line width=1pt]{\tikz[baseline]{\node[draw=ForestGreen!25, anchor=base, line width=1pt]{$#1$};}};}}

\begin{document}

\title{
  \Large\color{colD}\boldmath 
  Six-loop anomalous dimension of the $\phi^Q$ operator\\ in the $O(N)$
  symmetric model}

\author{\sc Alexander Bednyakov}\email{bednya@theor.jinr.ru} 
\author{\sc Andrey Pikelner}\email{pikelner@theor.jinr.ru} 
\affiliation{Bogoliubov Laboratory of Theoretical Physics, Joint Institute for
  Nuclear Research, 141980~Dubna, Russia}

\begin{abstract}
  A technique of large-charge expansion provides a novel opportunity for
  calculation of critical dimensions of operators $\phi^Q$ with fixed charge $Q$.
  In the small-coupling regime the polynomial structure of the anomalous
  dimensions can be fixed from a number of direct perturbative calculations for a
  fixed $Q$.
  At the six-loop level one needs to include new diagrams that correspond to
  operators with five or more legs. The latter never appeared before in
  scalar-theory calculations.
  Here we show how to compute the anomalous dimension of the operator
  $\phi^{Q=5}$ at the six-loop order.
  In combination with results for operators with $Q<5$, which are extracted from the six-loop
  beta-functions for general scalar theory, and with predictions from the large-charge expansion,    our calculation allows us to derive the answer for general-$Q$ anomalous dimensions.
  At the critical point resummation in three dimensions enables us to compare
  the critical exponents with results of Monte-Carlo simulations and large-$N$
  predictions.
\end{abstract}

\maketitle

\section{Introduction}
\label{sec:intro}

The renormalization group (RG) method allows one to systematically improve the accuracy of calculations in perturbation theory. 
The key objects of the method are the renormalization group functions, which specify the response of various quantities to a scale variation. 
Among many applications of the field-theoretical RG are studies of universal critical behaviour of different physical systems near second-order phase transitions. 
The well-known examples are three-dimensional $O(N)$ universality classes. The
corresponding critical indices can be derived by considering different operators
within the $\phi^4$ model in $d=4-2\ep$ dimensions. In particular, the
exponents are computed from the RG functions evaluated at a non-trivial
Wilson-Fisher \cite{Wilson:1971dc} (WF) fixed point, for which the quartic
coupling $g = g^* = \mathcal{O}(\ep)$. Recent six- \cite{Kompaniets:2017yct} and
even seven-loop \cite{Schnetz:2016fhy} results together with modern resummation
techniques (see \cite{Kompaniets:2017yct} and references therein) give very
precise critical exponents, which can be compared both to experimental
measurements and to values obtained by other methods.
While the expansion in small couplings seems natural and convenient, there exist
several non-pertubative approaches to the calculation of critical indices. For
example, in $O(N)$ theory one can consider the limit of large-$N$ and
systematically obtain corrections for scaling exponents as series in $1/N$
\cite{Vasiliev:1981dg,Vasiliev:1982dc}. The latter are valid for any space
dimensions and effectively resum infinite number of terms in the
$\ep$-expansion. The case of $\phi^Q$ operators was considered in
Ref.~\cite{Derkachov:1997ch}.
Recently, a different method was proposed \cite{Badel:2019oxl,Antipin:2020abu}
allowing one to study the anomalous dimensions of operators $\phi^Q$ with total
charge $Q$ semi-classically via operator-state correspondence.
In Ref.~\cite{Badel:2019oxl} the $U(1) = O(2)$ model is considered and a new 't
Hooft-like coupling is introduced $g^*Q$. The scaling dimension of
$\phi^Q$-type operators are written as
\begin{align}
	\Delta_{Q} & = \sum\limits_{i=-1}^{\infty} (g^*)^i \Delta_{i}(g^* Q) 
	= \frac{\Delta_{-1}(g^* Q)}{g^*} + \Delta_0(g^* Q) + \ldots
	\label{eq:deltaQ-semiclassical}
\end{align}
and the first two terms of the expansion in small $g^*$ for a fixed $g^*Q$ are
computed. The approach was generalized to the case of $O(N)$ model
\cite{Antipin:2020abu} together with $U(N) \times U(N)$ \cite{Antipin:2020rdw}
and $U(N) \times U(M)$ \cite{Antipin:2021akb} theories.
Expanding $\Delta_{-1}$ and $\Delta_{0}$ in small $g^*Q$, one can predict
leading and subleading terms in large $Q$ at arbitrary high loop. The latter can
be compared with existing perturbative calculations (see, e.g.,
Ref.~\cite{Jack:2021ypd,Jack:2021lja}), which for the case of the $O(N)$ model
are available (for arbitrary $Q$) up to five loops from recent paper
\cite{Jin:2022nqq}.
In this work, we use the results of Ref.~\cite{Antipin:2020abu} as an input
together with explicit computations for $Q=1...5$, to deduce the six-loop terms
in $\Delta_{Q}$ for general $Q$ in the $O(N)$ model. While the cases up to $Q=4$
can easily be treated given our general formulae for beta-functions valid in
arbitrary $\phi^4$ theory\cite{Bednyakov:2021ojn}, the computation of the
anomalous dimension for $Q=5$ constitutes the main technical challenge of the
current study.
The paper is organized as follows. In Section~\ref{sec:on-model} we briefly
review the $O(N)$ model and operators of our interest. In
Section~\ref{sec:calc-details} we discuss the details of calculation.
Section~\ref{sec:results} is devoted to our main results: six-loop expressions
for the anomalous and scaling dimension of $\phi^Q$-type operators in $O(N)$
model. Discussion and conclusion and can be found in Section~\ref{sec:concl}.

\section{Fixed-charge operators in $O(N)$ model}

The Euclidean Lagrangian that describes the $O(N)$-symmetric model is given by
\label{sec:on-model}
\begin{align}
	{\ensuremath{\mathcal{L}}} = \frac{1}{2} \left( \partial_{\mu} \vec{\phi} \cdot \partial_{\mu} \vec{\phi}    \right)
	+ \frac{g}{4!} \left(\vec\phi \cdot \vec \phi\right)^2
	,
	\label{eq:Lag_On}
\end{align}
where $\vec\phi = \{\phi_a\}$, $a=1...N$ is a $N$-component scalar field.
Following Ref.~\cite{Antipin:2020abu}, we consider lowest-lying $O(N)$ operators
of fixed total charge $Q$, which can be represented as
\begin{align}
	\phi^Q = d_{i_1....i_Q} 
		\phi_{i_1}
		\ldots
		\phi_{i_Q}, \quad d_{\ldots i \ldots i \ldots} = 0
		\label{eq:fixed-Q-operators}
\end{align}
with $d_{i_1 \ldots i_Q}$ being symmetric traceless tensor. 
	To compute the anomalous dimension $\gamma_Q$ of \eqref{eq:fixed-Q-operators}
perturbatively one can consider the insertion of the $\phi^Q$ operator in the
one-particle irreducible (1PI) Green function with $Q$ external legs (see.,
e.g., \cite{Jack:2021lja,Jin:2022nqq} for four- and five-loop results). Using
dimensional regularization \cite{tHooft:1973mfk} with $d=4 - 2 \ep$ and modified
minimal subtraction scheme (\MS), one extracts the renormalization constants
$Z_{\phi^Q}$, which relate bare fields $\phi_B$ to the renormalized operator
$[\phi^Q]$ :
\begin{align}
	\phi_B^Q = Z_{\phi^Q} [\phi^Q].
\label{eq:ZQ-definition}
\end{align}
The anomalous dimension can be cast in the following general form:
\begin{align}
  \label{eq:gammaQ-res}
  \gamma_Q(g) \equiv \frac{\partial \log Z_{\phi^Q}}{\partial g} (- 2 \epsilon g + \beta_g)  & = 
  Q \sum \limits_{l=1}^{\infty} g^l \gamma_Q^{(l)}, \qquad \gamma_Q^{(l)}  = 
  \sum \limits_{r=0}^{l} 
  \sum \limits_{s=0}^{l-r} 
  Q^r N^s \gamma^{(l)}_{r,s}, \quad \gamma^{(l)}_{0,l}=0.
\end{align}
Here $\beta_g$ is 4d part of the well-known beta function of the self-coupling
$g$ known up to six loops from Ref.~\cite{Kompaniets:2017yct}. In
Eq.~\eqref{eq:gammaQ-res} we sum over $l$-loop contributions $\gamma_Q^{(l)}$.
The latter are polynomials in $Q$ up to degree $l$, and for each monomial $Q^r$
the coefficient is a polynomial in $N$ up to degree $l-r$. The coefficients
$\gamma^{(l)}_{r,s}$ are just numbers.
Let us mention here that $Q=1$ case corresponds to the field anomalous dimension
\cite{Kompaniets:2017yct}, while, due to tensor nature of the operator, $Q=2$ is
related to the crossover exponent\footnote{And not to the anomalous dimension of
the singlet $\phi^2 \equiv \phi_i^2$.}\cite{Kompaniets:2019zes}.
Evaluating $\gamma_Q$ at the WF fixed point $g=g^*\simeq \frac{6 \ep}{N+8}$, we
compute the scaling dimensions of the operators in the form of $\ep$-expansion:
\begin{align}
  \label{eq:deltaQ-def}
  \Delta_Q & = Q (1 - \ep) + \gamma_{Q}(g^*) = Q + Q \sum_{l=1}^\infty \sum_{r=0}^{l} \sum_{p=r}^{2l-r}(2 \ep)^l \frac{Q^r}{(N+8)^p} P_{r,p}^{(l)}, \quad P_{0,0}^{(l)} = 0, ~P_{0,2l}^{(l)} = 0.
\end{align}
The ultimate goal of this paper is to compute 6-loop contributions
$\gamma^{(6)}_{r,s}$ and $P^{(6)}_{r,p}$ to Eqs.~\eqref{eq:gammaQ-res} and
\eqref{eq:deltaQ-def}, respectively. In what follows, we briefly review our
approach to the calculation.

\section{Calculation details}
\label{sec:calc-details}

\begin{figure}[t]
  \includegraphics[width=0.9\textwidth]{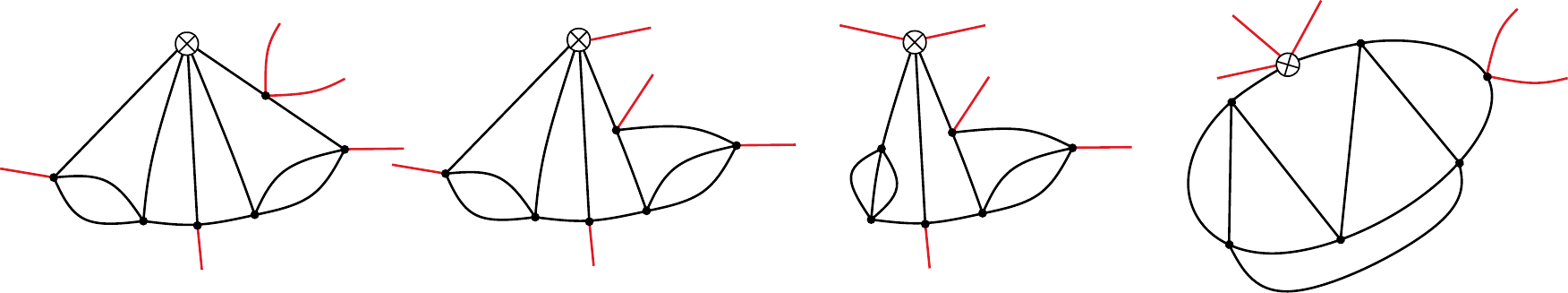}
  \caption{Examples of non-factorizable 6-loop diagrams contributing to 1PI
    five-point function with an operator insertion (denoted by cross). Only
    integrals similar to the first one require special treatment. All the others are
    known from the $\phi^4$ renormalization \cite{Kompaniets:2017yct}.}
  \label{fig:non-factorizable}
\end{figure}

According to Eq.~\eqref{eq:gammaQ-res}, six-loop contribution to the anomalous
dimension is a polynomial in $Q$
\begin{align}
	\gamma_Q^{(6)} = & 
	Q^0
	\Bigl(
		    \gamma^{(6)}_{0,0} + 
		N   \gamma^{(6)}_{0,1}  + 
		N^2 \gamma^{(6)}_{0,2}  + 
		N^3 \gamma^{(6)}_{0,3}  + 
		N^4 \boxN{\gamma^{(6)}_{0,4}}  + 
		N^5 \boxN{\gamma^{(6)}_{0,5}}  
	\Bigr) \nonumber\\
	 + & Q^1 \Bigl(
		    \gamma^{(6)}_{1,0} + 
		N   \gamma^{(6)}_{1,1}  + 
		N^2 \gamma^{(6)}_{1,2}  + 
		N^3 \gamma^{(6)}_{1,3}  + 
		N^4 \boxN{\gamma^{(6)}_{1,4}}  + 
		N^5 \boxNJ{\gamma^{(6)}_{1,5}}
	\Bigr) \nonumber\\
	 + & Q^2 \Bigl(
		    \gamma^{(6)}_{2,0} + 
		N   \gamma^{(6)}_{2,1}  + 
		N^2 \gamma^{(6)}_{2,2}  + 
		N^3 \gamma^{(6)}_{2,3}  + 
		N^4 \boxNJ{\gamma^{(6)}_{2,4}}
	\Bigr) \nonumber\\
	 + & Q^3 \Bigl(
		    \gamma^{(6)}_{3,0} + 
		N   \gamma^{(6)}_{3,1}  + 
		N^2 \gamma^{(6)}_{3,2}  + 
		N^3 \boxJ{\gamma^{(6)}_{3,3}}  
	\Bigr) \nonumber\\
	 + & Q^4 \Bigl(
		    \gamma^{(6)}_{4,0} + 
		N   \gamma^{(6)}_{4,1}  + 
		N^2 \boxJ{\gamma^{(6)}_{4,2}}   
	\Bigr) \nonumber\\
	 + & Q^5 \Bigl(
       \boxL{\gamma^{(6)}_{5,0}} + 
		N   \boxLJ{\gamma^{(6)}_{5,1}}   
       \Bigr) \nonumber\\
	 + & Q^6 
       \boxLJ{\gamma^{(6)}_{6,0}}
       ,
\label{eq:gammaQ_6L}
\end{align}
which has seven $N$-dependent coefficients. The latter can, in principle, be
determined from explicit perturbative results for seven anomalous dimensions
corresponding, e.g., to $Q=1...7$. Frames of different color
in~\eqref{eq:gammaQ_6L} highlight contributions known from other methods and are
discussed in detail in Section~\ref{sec:results}.
In this paper, we compute the anomalous dimensions for all operators up to
$Q=5$, giving five independent constraints on \eqref{eq:gammaQ_6L}. The
remaining two constraints are derived from the semi-classical result
\eqref{eq:deltaQ-semiclassical} of Ref.~\cite{Antipin:2020abu} expanded in $g^*
Q$ up to relevant order\footnote{We fix a small misprint $2/279\to 2/729$ in the
published version of Ref.~\cite{Antipin:2020abu}}:
\begin{align}
	\text{6-loop}:~~& \left( 
                    -\frac{572}{243} Q + \frac{2}{729} [
                    10191- 64 N - 2 \zeta_3 (1327 + 160 N)  - 2 \zeta_5 (1441 + 80 N)  \right. \nonumber\\
                  & \left. \vphantom{\frac{572}{243}} - 70 \zeta_7( 46 + N) - 21 \zeta_9 (126 + N) ]\right) (g^*Q)^6.
                    \label{eq:Antipin-6L}
\end{align}
As it was noted in Ref.~\cite{Jack:2021lja}, the contribution
\eqref{eq:Antipin-6L} to the anomalous dimension $\gamma_Q(g^*)$ derived
initially under the assumption $g=g^*$ is also valid away from the fixed point,
so we can immediately read off the coefficients $\gamma^{(6)}_{6,0}$,
$\gamma^{(6)}_{5,0}$, and $\gamma^{(6)}_{5,1}$ from Eq.~\eqref{eq:Antipin-6L}.
It turns out that the calculation of the operators with charge up to $Q=4$ is
trivial. One can use our six-loop result \cite{Bednyakov:2021ojn} for the RG
functions in the most general renormalizable $\phi^4$ theory. For example, to
compute $\gamma_{Q=4}$ we introduce a coupling $\lambda_4$ for the $\phi^{Q=4}$
operator. To extract the corresponding anomalous dimension from the
beta-function $\beta_{abcd}$ of general self-coupling $\lambda_{abcd}$, we
substitute
\begin{equation}
	\lambda_{abcd} \rightarrow \frac{g}{3} \left( \delta_{ab} \delta_{cd} + \delta_{ac} \delta_{bd} + \delta_{ad} \delta_{bc}\right) + \lambda_4 d_{abcd}
\end{equation}
and keep only terms that are linear in $\lambda_4$. The beta-function of
$\lambda_4$ is related to $\gamma_{Q=4}$ defined in \eqref{eq:gammaQ-res} via
\begin{equation}
	\beta_{\lambda_4}(g) = \lambda_4 \gamma_{Q=4}(g).
\end{equation}
The cases $Q=2$ and $Q=3$ are treated in a similar fashion, i.e., we replace the
mass parameter and the trilinear coupling by traceless symmetric tensors with
two and three indices
\begin{equation}
	m^2_{ab} \to \lambda_2 d_{ab}, \qquad h_{abc} \to \lambda_3 d_{abc}.
\end{equation}
It is worth mentioning that we routinely utilize FORM \cite{Vermaseren:1992vn}
to implement traceless condition on $d_{i_1...i_Q}$ \eqref{eq:fixed-Q-operators}
and to contract dummy indices.
The case $Q=5$ deserves special attention. We use DIANA \cite{Tentyukov:1999is}
to generate five-point 1PI Green functions with a $\phi^{Q=5}$ insertion. The
corresponding Feynman rule again involves the traceless symmetric tensor $d_{abcde}$.
After carrying out $O(N)$ algebra and factoring $d_{abcde}$, we are left with
scalar integrals multiplied by polynomials in $N$. It is obvious that some of
the indices entering $d_{i_1...i_5}$ can be external (see, e.g.,
Fig.~\ref{fig:non-factorizable}). In this case we effectively have a 1PI loop
diagram with reduced number of external legs.

\begin{figure}[t]
  \includegraphics[width=0.6\textwidth]{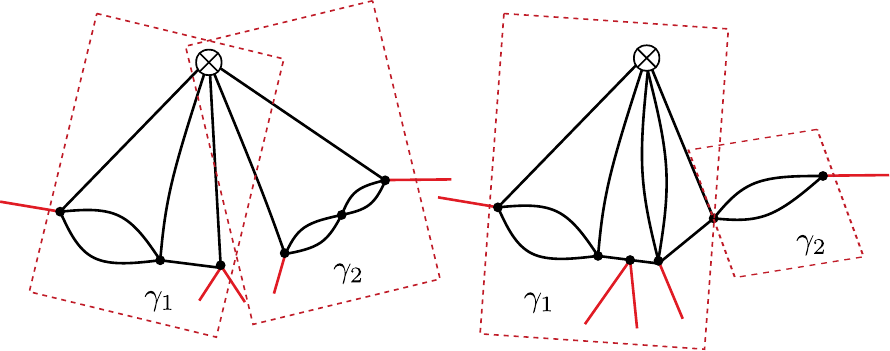}
  \caption{Examples of factorizable diagrams for which $\KRP (G) = \KRP (\gamma_1) \cdot \KRP (\gamma_2)$.}
  \label{fig:factorizable}
\end{figure}

To extract the renormalization constant $Z_5$ from the ultraviolet (UV)
divergences, we apply $\KRP$-operation to each logarithmically-divergent diagram
$G_i$:
\begin{equation}
	Z_\phi^5 Z_5 = 1 - \sum_i \KRP G_i,
\end{equation}
where $Z_\phi$ is a field renormalization constant $\phi_B = Z_\phi \phi$.
The $\KRP$-operation can be written recursively as 
\begin{equation}
  \KRP G  
  = \mathcal{K} G + \sum\limits_{\{\gamma\}} \mathcal{K}\big[\prod\limits_{\gamma_i \in \{\gamma\}}(- \KRP \gamma_i) \ast G/\{\gamma\} \big],
  \label{eq:KRp}
\end{equation}
where $G$ is the original diagram, and $\mathcal{K} G$ extracts its singular
$\mathcal{O}(1/\ep)$ part. The sum goes over all sets of disjoint UV-divergent
1PI subgraphs $\{\gamma\} = \cup_i \gamma_i$ (with $G$ itself excluded), and
$G/\{\gamma\}$ is a co-graph obtained from $G$ after shrinking all $\gamma_i$
belonging to $\{\gamma\}$. To implement \eqref{eq:KRp} at the six-loop order,
one needs to compute all lower-loop counterterms $\KRP \gamma_i$. Some of the
diagrams contain cut-vertices (see, e.g, Fig.~\ref{fig:factorizable}). In this
case $\KRP$ factorizes. In spite of the fact that these diagrams do not
contribute to the anomalous dimension of the operator, we keep them for further
crosschecks (see below).
Drastic simplification comes from the application of the infrared (IR)
rearrangement trick \cite{Vladimirov:1979zm} to the logarithmically divergent
diagrams. One can set all but one of the external momenta to zero and re-route
the momentum flow in a way to avoid (as much as possible) the appearance of
spurious IR divergences. The choice of the IR-safe routing together with the
UV-subgraph identification was automated by means of the private computer code.
It turns out that all integrals but one entering $\mathcal{K} G$ and
$G/\{\gamma\}$ can be calculated with IR-safe non-exceptional external momentum
routing in terms of graphical functions
\cite{Schnetz:2013hqa,Schnetz:2016fhy,Borinsky:2021gkd} implemented in
\texttt{HyperlogProcedures} package\footnote{Available for download from
\url{https://www.math.fau.de/person/oliver-schnetz/}}.
There remains a single diagram for which the IR-safe routing leads to an
integral not calculable with this approach. Due to this, the external momentum
routing was changed at the price of introduction of IR divergent subgraphs. The
latter were treated manually via infrared $\KRS$ operation
\cite{Chetyrkin:1982nn,Chetyrkin:2017ppe,doi:10.1142/4733,Herzog:2017bjx}. Further details are
provided in Appendix~\ref{sec:KRstar-details}.
Given $Z_5$, the required anomalous dimension is derived via
\begin{align}
	\gamma_{\!Q=5} = - \frac{\partial\log{Z_5}}{\partial \log{\mu}} = ( 2 \epsilon g - \beta_g) \frac{\partial \log{Z_5}}{\partial g}.
	\label{eq:gamma_Q5}
\end{align}
As usual only single poles in $\ep$ contribute to $\gamma_{Q=5}$. However, the
crucial crosscheck of the obtained expression is the cancellations of $\ep$
poles in the final formula for \eqref{eq:gamma_Q5}.

\section{Results}
\label{sec:results}

Given large-charge prediction \eqref{eq:deltaQ-semiclassical} and explicit
results for the anomalous dimensions of the first five operators $\phi^Q$, we
can fix all the coefficients in $\gamma_Q^{(6)}$:
\begin{equation}
  \gamma_Q^{(6)}  = 
  \sum \limits_{r=0}^{6} 
  \sum \limits_{s=0}^{6-r} 
  Q^r N^s \gamma^{(6)}_{r,s},
  \quad
  \gamma^{(6)}_{0,6}=0,
  \label{eq:gam6q-struct}
\end{equation}
as\footnote{Six-loop results for $\gamma_Q$ and $\Delta_Q$ are available
  as ancillary files with the arXiv version of the paper}
{\allowdisplaybreaks
  \begin{align}
    \boxLJ{\gamma^{(6)}_{6, 0}} & =   -\frac{572}{243},\\
    \boxLJ{\gamma^{(6)}_{5, 1}} & =   -\frac{640 \zeta_{3}}{729}-\frac{320 \zeta_{5}}{729}-\frac{140 \zeta_{7}}{729}-\frac{14 \zeta_{9}}{243}-\frac{128}{729} ,\\
    \boxL{\gamma^{(6)}_{5, 0}} & =   -\frac{5308 \zeta_{3}}{729}-\frac{5764 \zeta_{5}}{729}-\frac{6440 \zeta_{7}}{729}-\frac{196 \zeta_{9}}{27}+\frac{6794}{243},\\
    \boxJ{\gamma^{(6)}_{4, 2}} & =   -\frac{70 \zeta_{3}^2}{729}+\zeta_{3} \left(-\frac{46 \zeta_{5}}{729}-\frac{14}{81}\right)-\frac{100 \zeta_{5}}{729}-\frac{49 \zeta_{7}}{729}+\frac{14}{243}+\frac{7 \pi ^4}{7290}+\frac{10 \pi ^6}{137781}+\frac{7 \pi ^8}{1968300},\\
    \gamma^{(6)}_{4, 1} & =   -\frac{236 \zeta_{3}^2}{243}-\frac{176 \zeta_{3}^3}{729}+\zeta_{3} \left(\frac{1564}{243}-\frac{808 \zeta_{5}}{729}\right)-\frac{490 \zeta_{5}}{243}-\frac{11935 \zeta_{7}}{1458} \nonumber\\
                                & -\frac{42412 \zeta_{9}}{6561}+\frac{932}{729}+\frac{182 \pi ^4}{10935}+\frac{100 \pi ^6}{45927}+\frac{49 \pi ^8}{196830},\\
    \gamma^{(6)}_{4, 0} & =   \frac{2368 \zeta_{3}^2}{729}-\frac{2720 \zeta_{3}^3}{729}+\zeta_{3} \left(\frac{39340}{729}-\frac{9208 \zeta_{5}}{729}\right)+\frac{26564 \zeta_{5}}{729}+\frac{10451 \zeta_{7}}{729} \nonumber\\
                                & -\frac{81448 \zeta_{9}}{6561}-\frac{102694}{729}+\frac{784 \pi ^4}{10935}+\frac{1760 \pi ^6}{137781}+\frac{868 \pi ^8}{492075},\\
    \boxJ{\gamma^{(6)}_{3, 3}} & =   \pi ^4 \left(\frac{13 \zeta_{3}}{87480}+\frac{1}{4860}\right)+\frac{2 \zeta_{3}}{243}-\frac{65 \zeta_{3}^2}{2916}-\frac{59 \zeta_{5}}{2916}-\frac{25 \zeta_{7}}{2916}-\frac{1}{243}+\frac{25 \pi ^6}{1102248},\\
    \gamma^{(6)}_{3, 2} & =   \pi ^4 \left(\frac{131 \zeta_{3}}{43740}-\frac{133}{87480}\right)+\frac{607 \zeta_{3}^2}{1458}-\frac{8 \zeta_{3}^3}{81}+\zeta_{3} \left(\frac{2537}{2916}-\frac{218 \zeta_{5}}{729}\right)-\frac{37 \zeta_{5}}{54} \nonumber\\
                                &-\frac{20143 \zeta_{7}}{5832}-\frac{1063 \zeta_{9}}{729}-\frac{\zeta_{5,3}}{3}-\frac{335}{729}+\frac{160 \pi ^6}{137781}+\frac{5417 \pi ^8}{22044960},\\
    \gamma^{(6)}_{3, 1} & =   \pi ^4 \left(\frac{296 \zeta_{3}}{10935}-\frac{49}{729}\right)-\frac{788 \zeta_{3}^2}{729}-\frac{128 \zeta_{3}^3}{81}+\zeta_{3} \left(-\frac{12296 \zeta_{5}}{729}-\frac{35633}{1458}\right) \nonumber\\
                                & +\frac{512 \zeta_{5}}{243}+\frac{11 \zeta_{7}}{2}-\frac{5738 \zeta_{9}}{729}-\frac{2 (7191 \zeta_{5,3}+3275)}{1215} \nonumber\\
                                & +\frac{860 \pi ^6}{137781}+\frac{1213171 \pi ^8}{275562000},\\
    \gamma^{(6)}_{3, 0} & =   \pi ^4 \left(\frac{368 \zeta_{3}}{3645}-\frac{7151}{21870}\right)-\frac{84580 \zeta_{3}^2}{729}+\frac{5152 \zeta_{3}^3}{729}+\zeta_{3} \left(-\frac{59144 \zeta_{5}}{729}-\frac{44186}{243}\right) \nonumber\\
                                & -\frac{59306 \zeta_{5}}{729}-\frac{12529 \zeta_{7}}{486}+\frac{46256 \zeta_{9}}{6561}-\frac{10768 \zeta_{5,3}}{135} \nonumber\\
                                & +\frac{91750}{243}-\frac{2585 \pi ^6}{137781}+\frac{727847 \pi ^8}{34445250},\\
    \boxNJ{\gamma^{(6)}_{2, 4}} & =   \frac{\zeta_{3}}{972}+\frac{\zeta_{3}^2}{729}-\frac{\zeta_{5}}{486}+\frac{1}{5832}-\frac{\pi ^4}{174960}+\frac{\pi ^6}{1102248},\\
    \gamma^{(6)}_{2, 3} & =   \pi ^4 \left(-\frac{53 \zeta_{3}}{43740}-\frac{1}{9720}\right)+\frac{7 \zeta_{3}^2}{36}+\zeta_{3} \left(\frac{5 \zeta_{5}}{729}-\frac{631}{5832}\right)-\frac{35 \zeta_{5}}{1944}\nonumber\\
                                &-\frac{7735 \zeta_{7}}{11664}-\frac{\zeta_{5,3}}{45} +\frac{1619}{46656}+\frac{263 \pi ^6}{1102248}+\frac{2063 \pi ^8}{61236000},\\
    \gamma^{(6)}_{2, 2} & =   \pi ^4 \left(\frac{641}{43740}-\frac{449 \zeta_{3}}{21870}\right)-\frac{163 \zeta_{3}^2}{729}+\frac{32 \zeta_{3}^3}{729}-\frac{283 \zeta_{5}}{162}+\zeta_{3} \left(\frac{448 \zeta_{5}}{243}-\frac{3535}{972}\right)\nonumber\\
                                & +\frac{553 \zeta_{7}}{5832} -\frac{22964 \zeta_{9}}{6561}+\frac{4 \zeta_{5,3}}{45}+\frac{3541}{3888}+\frac{241 \pi ^6}{137781}+\frac{29453 \pi ^8}{137781000},\\
    \gamma^{(6)}_{2, 1} & =   \pi ^4 \left(\frac{5599}{21870}-\frac{1426 \zeta_{3}}{10935}\right)+\frac{12799 \zeta_{3}^2}{729}+\frac{2992 \zeta_{3}^3}{729}-\frac{43871 \zeta_{5}}{1458}+\zeta_{3} \left(\frac{42052 \zeta_{5}}{729}+\frac{14243}{486}\right) \nonumber\\
                                & -\frac{46361 \zeta_{7}}{972}-\frac{142852 \zeta_{9}}{6561}+\frac{14428 \zeta_{5,3}}{405}+\frac{41047}{5832}+\frac{185 \pi ^6}{275562}-\frac{1274101 \pi ^8}{137781000},\\
    \gamma^{(6)}_{2, 0} & =   \pi ^4 \left(\frac{3449}{3645}-\frac{3824 \zeta_{3}}{10935}\right)+\frac{380672 \zeta_{3}^2}{729}+\frac{32 \zeta_{3}^3}{729}-\frac{5050 \zeta_{5}}{81}+\zeta_{3} \left(\frac{100720 \zeta_{5}}{243}+\frac{22307}{81}\right) \nonumber\\
                                & -\frac{320719 \zeta_{7}}{1458}-\frac{596264 \zeta_{9}}{6561}+\frac{26944 \zeta_{5,3}}{81}-\frac{3367853}{5832}+\frac{8146 \pi ^6}{137781}-\frac{299533 \pi ^8}{3444525},\\
    \boxNJ{\gamma^{(6)}_{1, 5}} & =   -\frac{\zeta_{3}}{11664}+\frac{\zeta_{5}}{3888}-\frac{1}{23328}-\frac{\pi ^4}{699840},\\
    \boxN{\gamma^{(6)}_{1, 4}} & =   \frac{169 \zeta_{3}}{29160}-\frac{\zeta_{3}^2}{243}+\frac{13 \zeta_{5}}{1458}-\frac{299}{103680}-\frac{37 \pi ^4}{874800}-\frac{\pi ^6}{367416},\\
    \gamma^{(6)}_{1, 3} & =   \pi ^4 \left(\frac{1091 \zeta_{3}}{437400}-\frac{493}{437400}\right)-\frac{659 \zeta_{3}^2}{1620}+\zeta_{3} \left(\frac{12001}{58320}-\frac{5 \zeta_{5}}{243}\right)+\frac{3389 \zeta_{5}}{14580} \nonumber\\
                                & +\frac{1636 \zeta_{7}}{729}+\frac{\zeta_{5,3}}{15}-\frac{10403}{155520}-\frac{137 \pi ^6}{122472}-\frac{2063 \pi ^8}{20412000},\\
    \gamma^{(6)}_{1, 2} & =   \pi ^4 \left(\frac{7837 \zeta_{3}}{218700}-\frac{539}{9720}\right)-\frac{4834 \zeta_{3}^2}{3645}+\frac{136 \zeta_{3}^3}{243}+\zeta_{3} \left(\frac{96511}{29160}-\frac{340 \zeta_{5}}{729}\right)\nonumber\\
                                & +\frac{48371 \zeta_{5}}{3645}+\frac{171947 \zeta_{7}}{5832}+\frac{45287 \zeta_{9}}{2187}+\frac{521 \zeta_{5,3}}{225} \nonumber\\
                                & -\frac{10499}{2160}-\frac{53213 \pi ^6}{2755620}-\frac{7693807 \pi ^8}{2755620000},\\
    \gamma^{(6)}_{1, 1} & =   \pi ^4 \left(\frac{8158 \zeta_{3}}{54675}-\frac{6853}{10935}\right)-\frac{135304 \zeta_{3}^2}{3645}+\frac{64 \zeta_{3}^3}{27}+\zeta_{3} \left(-\frac{8956 \zeta_{5}}{729}-\frac{256211}{14580}\right) \nonumber\\
                                & +\frac{424234 \zeta_{5}}{3645}+\frac{14111 \zeta_{7}}{54}+\frac{159772 \zeta_{9}}{729}-\frac{13942 \zeta_{5,3}}{675} \nonumber\\
                                & -\frac{4451899}{116640}-\frac{77492 \pi ^6}{688905}-\frac{16401281 \pi ^8}{1377810000},\\
    \gamma^{(6)}_{1, 0} & =   \pi ^4 \left(\frac{3484 \zeta_{3}}{18225}-\frac{97517}{54675}\right)-\frac{3006466 \zeta_{3}^2}{3645}+\frac{4640 \zeta_{3}^3}{729}+\zeta_{3} \left(-\frac{302656 \zeta_{5}}{729}-\frac{1567481}{7290}\right) \nonumber\\
                                & +\frac{314518 \zeta_{5}}{729}+\frac{6962111 \zeta_{7}}{7290}+\frac{4163188 \zeta_{9}}{6561}-\frac{290408 \zeta_{5,3}}{675} \nonumber\\
                                & +\frac{7820977}{19440}-\frac{215048 \pi ^6}{688905}+\frac{24913489 \pi ^8}{344452500},\\
    \boxN{\gamma^{(6)}_{0, 5}} & =   \frac{\zeta_{3}}{15552}-\frac{\zeta_{5}}{3888}+\frac{1}{248832}+\frac{\pi ^4}{466560},\\
    \boxN{\gamma^{(6)}_{0, 4}} & =   -\frac{89 \zeta_{3}}{12960}+\frac{2 \zeta_{3}^2}{729}-\frac{5 \zeta_{5}}{729}+\frac{733}{466560}+\frac{89 \pi ^4}{1749600}+\frac{\pi ^6}{551124},\\
    \gamma^{(6)}_{0, 3} & =   \pi ^4 \left(\frac{2207}{1749600}-\frac{313 \zeta_{3}}{218700}\right)+\frac{3401 \zeta_{3}^2}{14580}+\zeta_{3} \left(\frac{10 \zeta_{5}}{729}-\frac{1667}{19440}\right)-\frac{61 \zeta_{5}}{270} \nonumber\\
                                & -\frac{18341 \zeta_{7}}{11664}-\frac{2 \zeta_{5,3}}{45}+\frac{69623}{933120}+\frac{965 \pi ^6}{1102248}+\frac{2063 \pi ^8}{30618000},\\
    \gamma^{(6)}_{0, 2} & =   \pi ^4 \left(\frac{197}{4374}-\frac{667 \zeta_{3}}{36450}\right)+\frac{4334 \zeta_{3}^2}{3645}-\frac{368 \zeta_{3}^3}{729}+\zeta_{3} \left(-\frac{740 \zeta_{5}}{729}-\frac{2297}{9720}\right)-\frac{4519 \zeta_{5}}{405} \nonumber\\
                                & -\frac{16885 \zeta_{7}}{648}-\frac{103330 \zeta_{9}}{6561}-\frac{466 \zeta_{5,3}}{225}+\frac{100471}{19440}+\frac{11617 \pi ^6}{688905}+\frac{1069637 \pi ^8}{459270000},\\
    \gamma^{(6)}_{0, 1} & =   \pi ^4 \left(\frac{38221}{87480}-\frac{836 \zeta_{3}}{18225}\right)+\frac{78049 \zeta_{3}^2}{3645}-\frac{3392 \zeta_{3}^3}{729}+\zeta_{3} \left(\frac{112751}{14580}-\frac{6664 \zeta_{5}}{243}\right) \nonumber\\
                                & -\frac{320224 \zeta_{5}}{3645}-\frac{614519 \zeta_{7}}{2916}-\frac{1200664 \zeta_{9}}{6561}-\frac{6344 \zeta_{5,3}}{2025} \nonumber\\
                                & +\frac{9046223}{233280}+\frac{146159 \pi ^6}{1377810}+\frac{1894453 \pi ^8}{114817500},\\
    \gamma^{(6)}_{0, 0} & =   \pi ^4 \left(\frac{3148 \zeta_{3}}{54675}+\frac{242993}{218700}\right)+\frac{504412 \zeta_{3}^2}{1215}-\frac{2368 \zeta_{3}^3}{243}-\frac{232190 \zeta_{5}}{729} \nonumber\\
                                & +\zeta_{3} \left(\frac{68848 \zeta_{5}}{729}+\frac{550921}{7290}\right)-\frac{1736897 \zeta_{7}}{2430}-\frac{1161368 \zeta_{9}}{2187}+\frac{359144 \zeta_{5,3}}{2025} \nonumber\\
                                & -\frac{9723527}{116640}+\frac{8608 \pi ^6}{32805}-\frac{11713 \pi ^8}{1417500}.
  \end{align}
}
Evaluating the anomalous dimension at the fixed point, we obtain the
$\ep$-expansion of the scaling dimension $\Delta_Q$ \eqref{eq:deltaQ-def}. We
reproduce the five-loop results \cite{Jin:2022nqq}. The new six-loop
coefficients are given by
{\allowdisplaybreaks
  \begin{align}
    P_{6,6}^{(6)} & = -1716, \\
    P_{5,7}^{(6)} & = -38400, \\
    P_{5,6}^{(6)} & = -2 (94 \zeta_{3}+1602 \zeta_{5}+2660 \zeta_{7}+2478 \zeta_{9}-16463), \\
    P_{5,5}^{(6)} & = -2 (320 \zeta_{3}+160 \zeta_{5}+70 \zeta_{7}+21 \zeta_{9}+64), \\
    P_{4,8}^{(6)} & = -529200, \\
    P_{4,7}^{(6)} & = -24 (2212 \zeta_{3}+3500 \zeta_{5}+4725 \zeta_{7}-27264), \\
    P_{4,6}^{(6)} & = 3552 \zeta_{3}^2-1312 \zeta_{3}^3+51124 \zeta_{5}-4 \zeta_{3} (1422 \zeta_{5}+935) \nonumber \\
                & + 86975 \zeta_{7}+\frac{257848 \zeta_{9}}{9}-265526, \\
    P_{4,5}^{(6)} & = 412 \zeta_{3}^2-176 \zeta_{3}^3+1930 \zeta_{5}-24 \zeta_{3} (3 \zeta_{5}-437) \nonumber \\
                & -\frac{9107 \zeta_{7}}{2}-\frac{42412 \zeta_{9}}{9}+1898+\frac{14 \pi ^4}{15}+\frac{20 \pi ^6}{27}+\frac{7 \pi ^8}{50}, \\
    P_{4,4}^{(6)} & = -70 \zeta_{3}^2-100 \zeta_{5}-2 \zeta_{3} (23 \zeta_{5}+63)-49 \zeta_{7}+42+\frac{7 \pi ^4}{10}+\frac{10 \pi ^6}{189}+\frac{7 \pi ^8}{2700}, \\
    P_{3,9}^{(6)} & = -6048000, \\
    P_{3,8}^{(6)} & = -5040 (301 \zeta_{3}+275 \zeta_{5}-1602), \\
    P_{3,7}^{(6)} & = -79776 \zeta_{3}^2+\zeta_{3} (951996-95040 \zeta_{5})+1306920 \zeta_{5}+704025 \zeta_{7}-4540356, \\
    P_{3,6}^{(6)} & = -33870 \zeta_{3}^2+9760 \zeta_{3}^3-425010 \zeta_{5}+24 \zeta_{3} (1973 \zeta_{5}+856) -\frac{1029567 \zeta_{7}}{2}-\frac{152896 \zeta_{9}}{9} \nonumber \\
                & -\frac{23328 \zeta_{5,3}}{5} +1148614+\frac{386 \pi ^4}{5}+\frac{1100 \pi ^6}{63}+\frac{1044 \pi ^8}{875}, \\
    P_{3,5}^{(6)} & = -7984 \zeta_{3}^2+\frac{1}{60} \pi ^4 (336 \zeta_{3}-493)+17146 \zeta_{5}-\frac{3}{2} \zeta_{3} (5072 \zeta_{5}+44331) \nonumber \\
                & +\frac{286097 \zeta_{7}}{4}+11270 \zeta_{9}-\frac{23706 \zeta_{5,3}}{5}-10492-\frac{2150 \pi ^6}{189}+\frac{14419 \pi ^8}{42000}, \\
    P_{3,4}^{(6)} & = 986 \zeta_{3}^2-72 \zeta_{3}^3-\frac{1}{24} \pi ^4 (10 \zeta_{3}+221)+\zeta_{3} \left(\frac{5993}{4}-218 \zeta_{5}\right) \nonumber \\
                & +\frac{999 \zeta_{5}}{2}-\frac{18943 \zeta_{7}}{8}-1063 \zeta_{9}-243 \zeta_{5,3}-\frac{939}{2}+\frac{115 \pi ^6}{756}+\frac{5417 \pi ^8}{30240}, \\
    P_{3,3}^{(6)} & = \frac{1}{120} \pi ^4 (13 \zeta_{3}+18)+\frac{1}{4} \left(24 \zeta_{3}-65 \zeta_{3}^2-59 \zeta_{5}-25 \zeta_{7}-12\right)+\frac{25 \pi ^6}{1512}, \\
    P_{2,10}^{(6)} & = -68040000, \\
    P_{2,9}^{(6)} & = -604800 (51 \zeta_{3}-142), \\
    P_{2,8}^{(6)} & = -252 \left(-114928 \zeta_{3}+9072 \zeta_{3}^2-18200 \zeta_{5}+195133\right), \\
    P_{2,7}^{(6)} & = 6 \left(219240 \zeta_{3}^2+21 \zeta_{3} (3360 \zeta_{5}-72017)+351 \pi ^4\right) \nonumber\\
                &-30 (187650 \zeta_{5}+33957 \zeta_{7}-538628) , \\
    P_{2,6}^{(6)} & = 74436 \zeta_{3}^2-21856 \zeta_{3}^3+\frac{1}{5} \pi ^4 (648 \zeta_{3}-6619)+\zeta_{3} (447674-176400 \zeta_{5}) -\frac{760 \pi ^6}{9}\nonumber \\
                & +2091276 \zeta_{5}+\frac{1646491 \zeta_{7}}{2}-\frac{923144 \zeta_{9}}{9}+\frac{235872 \zeta_{5,3}}{5}-2793137-\frac{1508 \pi ^8}{125}, \\
    P_{2,5}^{(6)} & = 27831 \zeta_{3}^2+2480 \zeta_{3}^3-\frac{2}{15} \pi ^4 (297 \zeta_{3}-844) -\frac{476097 \zeta_{5}}{2}+\zeta_{3} \left(46708 \zeta_{5}+\frac{361793}{2}\right) \nonumber \\
                & -\frac{790853 \zeta_{7}}{4}+\frac{224572 \zeta_{9}}{9} + \frac{109116 \zeta_{5,3}}{5}+\frac{52581}{2}+\frac{24205 \pi ^6}{378}-\frac{40673 \pi ^8}{9000}, \\
    P_{2,4}^{(6)} & = -4162 \zeta_{3}^2+32 \zeta_{3}^3+\frac{7}{60} \pi ^4 (2 \zeta_{3}+351)+6263 \zeta_{5}+\zeta_{3} \left(1704 \zeta_{5}-\frac{11473}{2}\right) \nonumber \\
                & +\frac{133063 \zeta_{7}}{8}-\frac{22964 \zeta_{9}}{9}+\frac{2268 \zeta_{5,3}}{5}+\frac{14669}{8}-\frac{385 \pi ^6}{54}-\frac{11707 \pi ^8}{27000}, \\
    P_{2,3}^{(6)} & = \frac{1}{80} \left(8780 \zeta_{3}^2+6390 \zeta_{5}+100 \zeta_{3} (4 \zeta_{5}-63)-38675 \zeta_{7}-1296 \zeta_{5,3}+2175\right) \nonumber\\
                & -\frac{1}{120} \pi ^4 (106 \zeta_{3}+161)+\frac{11 \pi ^6}{72}+\frac{2063 \pi ^8}{84000}, \\
    P_{2,2}^{(6)} & = \frac{3 \zeta_{3}}{4}+\zeta_{3}^2-\frac{3 \zeta_{5}}{2}+\frac{1}{8}-\frac{\pi ^4}{240}+\frac{\pi ^6}{1512}, \\
    P_{1,11}^{(6)} & = -1020600000, \\
    P_{1,10}^{(6)} & = -27216000 (18 \zeta_{3}-41), \\
    P_{1,9}^{(6)} & = -7560 \left(-77856 \zeta_{3}+5184 \zeta_{3}^2+16800 \zeta_{5}+71741\right), \\
    P_{1,8}^{(6)} & = 126 \left(324144 \zeta_{3}^2+610860 \zeta_{5}+1250399+270 \pi ^4\right) \nonumber \\
                & -126 \left(30 \zeta_{3} (2016 \zeta_{5}+73393)+66150 \zeta_{7}\right), \\
    P_{1,7}^{(6)} & = -13852800 \zeta_{3}^2+50688 \zeta_{3}^3+\frac{1296}{5} \pi ^4 (9 \zeta_{3}-133)-6666360 \zeta_{5} \nonumber \\
                & +24 \zeta_{3} (110880 \zeta_{5}+2632463)-2469033 \zeta_{7}+966080 \zeta_{9}+\frac{497664 \zeta_{5,3}}{5}  \nonumber \\
                & -\frac{121888917}{4}+1200 \pi ^6-\frac{22272 \pi ^8}{875}, \\
    P_{1,6}^{(6)} & = \pi ^4 \left(\frac{24247}{2}-\frac{8172 \zeta_{3}}{5}\right)+1369356 \zeta_{3}^2-41696 \zeta_{3}^3-3917542 \zeta_{5}\nonumber \\
                & -\frac{3}{2} \zeta_{3} (155136 \zeta_{5}+4083031) +\frac{5677959 \zeta_{7}}{2}-\frac{8721100 \zeta_{9}}{9} \nonumber \\
                & -\frac{1313064 \zeta_{5,3}}{5}+\frac{28571673}{8}-\frac{24965 \pi ^6}{63}+\frac{782339 \pi ^8}{10500}, \\
    P_{1,5}^{(6)} & = -122928 \zeta_{3}^2+5696 \zeta_{3}^3+\frac{4}{15} \pi ^4 (951 \zeta_{3}-5767)+\zeta_{3} \left(\frac{249699}{4}-44604 \zeta_{5}\right)\nonumber \\
                & +878664 \zeta_{5} -\frac{1581769 \zeta_{7}}{4}+\frac{997516 \zeta_{9}}{9}-\frac{24138 \zeta_{5,3}}{5}-\frac{59181}{2}-\frac{1660 \pi ^6}{21}+\frac{68489 \pi ^8}{14000}, \\
    P_{1,4}^{(6)} & = \pi ^4 \left(\frac{61}{4}-\frac{35 \zeta_{3}}{12}\right)+\frac{21937 \zeta_{3}^2}{2}+664 \zeta_{3}^3+\zeta_{3} \left(\frac{831}{8}-2188 \zeta_{5}\right)-\frac{101553 \zeta_{5}}{2}-\frac{31659 \zeta_{7}}{8} \nonumber \\
                & +\frac{194069 \zeta_{9}}{9}+\frac{6237 \zeta_{5,3}}{5}-\frac{77953}{32}+\frac{9467 \pi ^6}{378}-\frac{147677 \pi ^8}{108000}, \\
    P_{1,3}^{(6)} & = -\frac{1415 \zeta_{3}^2}{4}+\frac{1}{120} \pi ^4 (247 \zeta_{3}-5)+\zeta_{3} \left(\frac{6225}{16}-15 \zeta_{5}\right)+\frac{2645 \zeta_{5}}{4}+1724 \zeta_{7}+\frac{243 \zeta_{5,3}}{5}\nonumber \\
                & -\frac{10273}{128} -\frac{1733 \pi ^6}{1512}-\frac{2063 \pi ^8}{28000}, \\
    P_{1,2}^{(6)} & = -4 \zeta_{3}-3 \zeta_{3}^2-\zeta_{5}-\frac{383}{256}+\frac{13 \pi ^4}{120}-\frac{\pi ^6}{504}, \\
    P_{1,1}^{(6)} & = \frac{1}{960} \left(-30 (2 \zeta_{3}-6 \zeta_{5}+1)-\pi ^4\right), \\
    P_{0,11}^{(6)} & = 1020600000, \\
    P_{0,10}^{(6)} & = 3402000 (144 \zeta_{3}-323), \\
    P_{0,9}^{(6)} & = 15120 \left(-37968 \zeta_{3}+2592 \zeta_{3}^2+8400 \zeta_{5}+33763\right), \\
    P_{0,8}^{(6)} & = -126 \left(309888 \zeta_{3}^2+661460 \zeta_{5}+1065945+270 \pi ^4\right) \nonumber \\
                & + 126 \left(
                  2 \zeta_{3} (30240 \zeta_{5}+1055743)+66150 \zeta_{7}
                  \right), \\
    P_{0,7}^{(6)} & = 12961872 \zeta_{3}^2-50688 \zeta_{3}^3-\frac{486\pi ^4}{5}  (24 \zeta_{3}-343)+12986880 \zeta_{5} \nonumber\\
                & -18 \zeta_{3} (166080 \zeta_{5}+3400493) +2778048 \zeta_{7}-966080 \zeta_{9}\nonumber \\
                & -\frac{497664 \zeta_{5,3}}{5}+\frac{87424317}{4}-1200 \pi ^6+\frac{22272 \pi ^8}{875}, \\
    P_{0,6}^{(6)} & = -1493358 \zeta_{3}^2+55104 \zeta_{3}^3+\frac{1}{10} \pi ^4 (15048 \zeta_{3}-116131)+1861686 \zeta_{5}\nonumber \\
                & +\zeta_{3} (367440 \zeta_{5}+6850142) -\frac{6511113 \zeta_{7}}{2} \nonumber \\
                & +\frac{3194632 \zeta_{9}}{3}+220104 \zeta_{5,3}-\frac{32994733}{16}+\frac{30445 \pi ^6}{63}-\frac{133639 \pi ^8}{2100}, \\
    P_{0,5}^{(6)} & = 109509 \zeta_{3}^2-8000 \zeta_{3}^3-\frac{1}{30} \pi ^4 (6588 \zeta_{3}-48821)-655038 \zeta_{5}+\zeta_{3} \left(5576 \zeta_{5}-\frac{608795}{2}\right) \nonumber \\
                & +\frac{2210257 \zeta_{7}}{4}-\frac{1280728 \zeta_{9}}{9}-\frac{61272 \zeta_{5,3}}{5}+\frac{892443}{32}+\frac{7055 \pi ^6}{378}-\frac{26969 \pi ^8}{31500}, \\
    P_{0,4}^{(6)} & = -\frac{15819 \zeta_{3}^2}{2}-624 \zeta_{3}^3+\frac{1}{80} \pi ^4 (248 \zeta_{3}-5521)+\frac{95231 \zeta_{5}}{2}+\zeta_{3} \left(748 \zeta_{5}+\frac{80693}{8}\right)\nonumber \\
                & -\frac{106765 \zeta_{7}}{8} -\frac{53846 \zeta_{9}}{3}-1458 \zeta_{5,3}+\frac{34005}{32}-\frac{4493 \pi ^6}{252}+\frac{8161 \pi ^8}{5040}, \\
    P_{0,3}^{(6)} & = \frac{1073 \zeta_{3}^2}{4}-\frac{7}{240} \pi ^4 (44 \zeta_{3}-73)-917 \zeta_{5}+\zeta_{3} \left(10 \zeta_{5}-\frac{3807}{8}\right)-\frac{19749 \zeta_{7}}{16}-\frac{162 \zeta_{5,3}}{5} \nonumber \\
                & +\frac{1693}{32}+\frac{173 \pi ^6}{168}+\frac{2063 \pi ^8}{42000}, \\
    P_{0,2}^{(6)} & = \frac{111 \zeta_{3}}{16}+2 \zeta_{3}^2+\frac{5 \zeta_{5}}{2}+\frac{343}{512}-\frac{59 \pi ^4}{480}+\frac{\pi ^6}{756}, \\
    P_{0,1}^{(6)} & = \frac{240 \zeta_{3}-960 \zeta_{5}+15+8 \pi ^4}{5120}.
  \end{align}
}

Provided results for coefficients $\gamma^{(l)}_{r,s}$ and $P_{r,p}^{(l)}$
in addition to the Riemann zeta functions $\zeta_n = \sum_{i=1}^{\infty} 1/i^n$
contain multiple zeta value $\zeta_{5,3} = \sum_{i=1}^{\infty}\sum_{j=1}^{i-1}
1/(i^5j^3) \approx 0.0377077$.
To verify our expressions we consider various limits of
Eq.~\eqref{eq:gammaQ-res}. For example, defining $J=Q/N$ we get an expansion in
$J$. Neglecting terms further suppressed by large $N$, we obtain (for $g=g^*$)
\begin{align}
	\frac{\Delta_Q}{Q} & \simeq  
	 1 - \ep   
 + \sum_{l=1}^{\infty} (6 \ep)^l \sum_{k=1}^{l} J^k \gamma^{(l)}_{k,l-k} 
	+ \mathcal{O}\left(\frac{1}{N} \right) \nonumber\\
			   & = 1 - \ep   
			   + \sum_{l=1}^{\infty} (2 \ep)^l \sum_{k=1}^{l} J^k P_{k,k}^{(l)}
	+ \mathcal{O}\left(\frac{1}{N} \right)
\label{eq:J-ep-expansion}
\end{align}
with
\begin{equation}
	P_{k,k}^{(l)} = 3^l \gamma^{(l)}_{k,l-k}.
\end{equation}
It turns out that the expansion of $\Delta_Q$ in small $J$ for large $N$ can be found for a general dimension $d$ \cite{Giombi:2020enj}:
\begin{equation}
	\frac{\Delta Q}{Q} = \frac{d}{2} - 1 + h_{2}(d) j + h_{3}(d) j^2 + \ldots, 
\label{eq:J-expansion}
\end{equation}
where, e.g.,
\begin{equation}
	h_2(d) = -\frac{2^{d-3} d \sin \left(\frac{\pi d}{2}\right) \Gamma \left( \frac{d-1}{2}\right)}{\pi^{3/2} \Gamma\left(\frac{d}{2} +1 \right)}.
\end{equation}
We follow \cite{Giombi:2020enj} to compute the $\ep$-expansion of $h_{i}(d)$,
for $i=2...7$, and find perfect agreement with our perturbative result
\eqref{eq:J-ep-expansion}. In this way, we check $\gamma^{(6)}_{r,6-r}$ for
$r=1...6$.
In addition, we also compare our result with the first two non-trivial orders of
large-$N$ expansion\cite{Derkachov:1997ch}, which begins as
\begin{equation}
	\frac{\Delta Q}{Q} = \frac{d}{2} - 1 + \frac{h_{2}(d)}{N} \left( Q - 2  + \frac{4}{d} \right)  +  \ldots, 
\label{eq:N-expansion}
\end{equation}
At six loops only five coefficients of Eq.~\eqref{eq:gammaQ-res} contribute to
large-$N$ expansion of our result \eqref{eq:deltaQ-def} up to
$\mathcal{O}(1/N^2)$. Two of them $\gamma^{(6)}_{1,5}$ and $\gamma^{(6)}_{2,4}$
also enter \eqref{eq:J-expansion}, while the comparison with $\ep$-expansion of
\eqref{eq:N-expansion} provides additional checks for $\gamma^{(6)}_{0,4}$,
$\gamma^{(6)}_{1,4}$, and $\gamma^{(6)}_{0,5}$.

\section{Discussion and Conclusion}
\label{sec:concl}

In this paper we derived the six-loop anomalous dimension of the charged
$\phi^Q$ operator in the $O(N)$ model. Our computation was based on the
combination of semi-classical results and explicit diagram calculations. Our
expression up to five loops coincides with that obtained recently in
Refs.~\cite{Henriksson:2022rnm,Jin:2022nqq}.
The utilized approach relies on diagram-by-diagram computation of Feynman
graphs with an $\phi^5$ insertion, and, thus, facilitates further six-loop studies of
the charged operators in models with other symmetries once
semi-classical results are available.
Given our result, one can apply various resummation techniques to the
$\ep$-expansion of $\Delta_Q$ and obtain numerical values of scaling dimension
in a bunch of three-dimensional $O(N)$ models.
%

%
%

\begin{table}[th]
	\centering
	\begin{tabular}{|c|c|c|c|c|c|c|c|c|c|c|}
		\hline
		\diagbox{$N$}{$Q$} & 1 & 2 & 3 & 4 &5 &6 &7 &8 &9 & 10  \\
		\hline
    2 &  0.5187(5) &1.2351(6)  & 2.1085(7)  & 3.112(2)  & 4.230(3)  & 5.45(1) & 6.75(2)   &8.15(2)   &9.62(3)   &11.17(3)  \\ \hline
    4 &  0.5181(5) & 1.1885(3) & 1.9856(14) & 2.892(3)  & 3.896(7)  & 4.99(2) & 6.16(3)   & 7.40(3)     & 8.72(4)     & 10.10(4)     \\ \hline
    6 &  0.5163(5) &1.1537(13) & 1.895(3)   & 2.728(5)  & 3.643(11) & 4.63(2) & 5.70(3)   &6.82(3)   &8.01(4)   &9.26(4)    \\ \hline
    8 &  0.5146(4) & 1.127(2)  & 1.827(5)   & 2.604(8)  & 3.45(2)   & 4.36(2) & 5.34(3)   & 6.37(3)  & 7.46(4)  & 8.60(4)     \\ \hline
	\end{tabular}
  \caption{Scaling dimensions of $\phi^Q$ ($d=3$) obtained by resumming six-loop result for $N=2,4,6,8$.}
  \label{tab:resummed_deltaQ}
\end{table}
Based on the fact that we are working at the same order of perturbation theory
and in the same theory, we use in straightforward manner the advanced technique of
Ref.~\cite{Kompaniets:2017yct} to compute numerical
values and uncertainties of $\Delta_Q$ for $Q=1\ldots 10$ and $N=2,4,6,8$ (see
Table~\ref{tab:resummed_deltaQ}). The latter can be compared to the Monte-Carlo results
\cite{Banerjee:2017fcx,Banerjee:2019jpw,Singh:2022akp}.
It is interesting to study how our computation matches the large $Q$ expansion
\cite{Hellerman:2015nra}
\begin{equation}
	\Delta_Q = c_{3/2} Q^{3/2} + c_{1/2} Q^{1/2} + c_0 + \mathcal{O}(Q^{-1/2})
  \label{eq:large-charge-3d}
\end{equation}
Here $c_{3/2}$, $c_{1/2}$ are $N$-dependent constants, while $c_0\approx
-0.0937254$ originates from Casimir energy and is universal. In
Ref.~\cite{Alvarez-Gaume:2019biu} the large-$N$ limit was considered and the
following predictions were obtained
\begin{align}
 	c_{3/2} = \frac{2}{3} N^{-1/2}, \quad c_{1/2} = \frac{1}{6} N^{1/2}
	\label{eq:AGpredictionC}
\end{align}
which are valid for $1\ll Q\ll N\ll Q^2$. 
\begin{table}[ht]
  \centering
  \begin{tabular}{|c||c|c|c|}
    \hline
    $N$ & $c_{3/2}$ & $c_{1/2}$ & $c_{-1/2}$ \\
    \hline\hline
    2   &  0.3324(6)     &  0.271(3)    &  0.009(3)  \\
    4   &  0.2925(4)     &  0.3241(13)  & -0.0048(11)\\
    6   &  0.2612(6)     &  0.371(3)    & -0.022(2) \\
    8   &  0.2371(9)     &  0.408(3)    & -0.036(3)\\
    \hline
  \end{tabular}
  \caption{Fit from resummation results using $Q=1\dots 6$ expressions with 
    $c_0 = -0.0937$ fixed and additional free parameter $c_{-1/2}$ in $c_{-1/2}
    Q^{-1/2}$ term appended to the ansatz~\eqref{eq:large-charge-3d}.}
  \label{tab:c0123fix}
\end{table}
\begin{figure}[th]
	\centering
  \includegraphics[width=\textwidth]{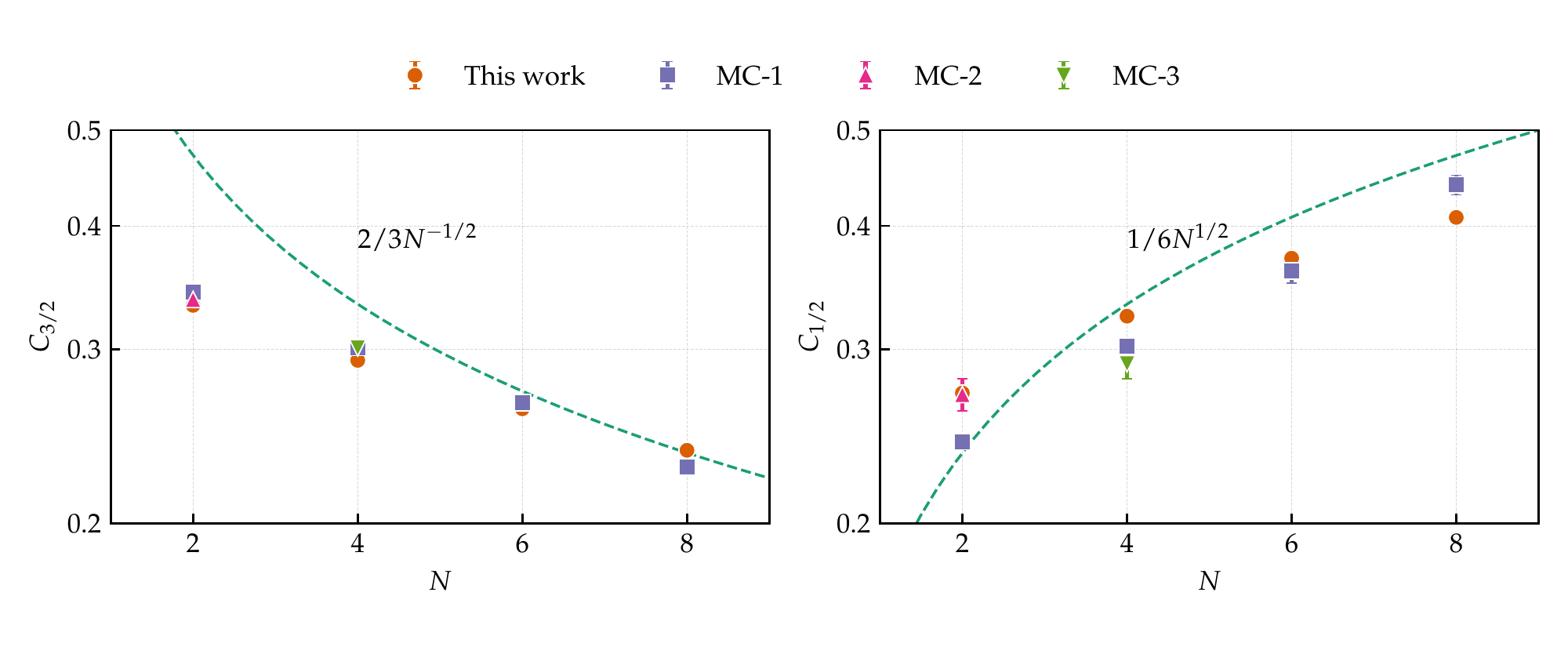}
	\caption{Leading coefficients in large-charge expansion $c_{3/2}$ and
    $c_{1/2}$ as functions of $N$. We also add results of Monte-Carlo simulation
    for $N=2,4,6,8$ from Ref.~\cite{Singh:2022akp} (MC-1), 
    for $N=2$ from Ref.~\cite{Banerjee:2017fcx}  (MC-2) and
    for $N=4$ from Ref.~\cite{Banerjee:2019jpw} (MC-3).}
	\label{fig:c32_c12_fit}
\end{figure}
We follow \cite{Banerjee:2017fcx,Singh:2022akp} and fit our numerical data to
\eqref{eq:large-charge-3d}. The extracted coefficient for a fixed $c_{0}$ are
given in Table~\ref{tab:c0123fix}. One can see satisfactory agreement between
our values and Monte-Carlo results
\cite{Banerjee:2017fcx,Banerjee:2019jpw,Singh:2022akp} (see
Fig.~\ref{fig:c32_c12_fit}). While our fits for $c_{3/2}$ are close to the
Monte-Carlo values and approaches large-$N$ limit \eqref{eq:AGpredictionC}, we
see a rather big discrepancies in $c_{1/2}$. We expect that our error estimates
according to Ref.~\cite{Kompaniets:2017yct} may be too optimistic and require
further studies. Nevertheless, we see that according to
Eq.~\eqref{eq:AGpredictionC} the coefficient $c_{3/2}$ decreases with $N$, while
$c_{1/2}$ increases.
We believe our result contributes important information to CFT data (see,
e.g.,\cite{Henriksson:2022rnm} for review). While we rely on non-perturbative
large $Q$ expressions to fix part of the coefficients in the anzats
\eqref{eq:gammaQ_6L}, it would be interesting to reproduce them independently,
e.g., by considering Green functions with $\phi^{Q=6}$ and $\phi^{Q=7}$
insertions. We postpone this verification for the future.

\acknowledgements
We thank O. Antipin, J. Henriksson, S. Kosvous, M.Kompaniets, A. Kudlis, and
N.Lebedev for fruitful discussions. Furthermore, we are grateful to the Joint
Institute for Nuclear Research for using their supercomputer ``Govorun.''
\appendix

\section{Details of calculation with $\KRS$ operation}
\label{sec:KRstar-details}
Here, we present the expression used for manual computation of a single diagram
by means of $\KRS$ operation. We closely follow the notation of
Refs.~\cite{Herzog:2017bjx,doi:10.1142/4733}. One can see that there is only one non-trivial
IR-divergent counterterm
\tikzset{
  IR/.style={line width=1pt,draw=red},
  prop/.style={line width=1pt,draw=colA},
  leg/.style={line width=1pt,draw=colD}}
\begin{align}
    \gamma_{\rm IR} = \left(
    \vcenter{\hbox{
    \begin{tikzpicture}[use Hobby shortcut, scale=1.0]
      \draw[IR] (0,0.45) -- (0,-0.45);
      \fill (0,0) circle (1.5pt);
      \draw (0,0.5) circle (1.5pt);
      \draw (0,-0.5) circle (1.5pt);
    \end{tikzpicture}       
    }}\right) = \frac{1}{\ep},
\end{align}
which cancels spurious IR divergences appearing in $\Gamma_i = G/\gamma_i$ for a
UV subgraph $\gamma_i$. The graph $\tilde \Gamma_i = \Gamma_i\setminus \gamma_{\rm IR}$
is obtained from $\Gamma_i$ by deleting lines and internal vertices (denoted by
filled dots) of $\gamma_{\rm IR}$. One can see that even if $\Gamma_i$ can be zero in
dimensional regularization the corresponding IR counterterm can give a
non-trivial contribution to the final result. All $\KRP(\gamma_i)$ are known
from lower-loop calculations, while $G$, $\tilde G$, $\Gamma_i$, and $\tilde
\Gamma_i$ calculated with \texttt{HyperlogProcedures}.
\begin{align}
  \label{eq:KRS-calc}
  & \KRS \vcenter{\hbox{
    \begin{tikzpicture}[use Hobby shortcut, scale=1.0]
      \draw[prop] (0,0) circle (1.0);
      \coordinate (v1) at (180:1.0);
      \coordinate (v2) at (225:1.0);
      \coordinate (v3) at (270:1.0);
      \coordinate (v4) at (0:1.0);
      \coordinate (v5) at (90:0.5);
      \coordinate (v6) at (270:0.5);
      \coordinate (v7) at (0.6,0.35);
      \coordinate (v7ul) at (-0.6,0.35);
      \coordinate (v7dl) at (-0.6,-0.35);
      \coordinate (v7dr) at (0.6,-0.35);
      \draw[prop] (v1) .. (v7ul) .. (v5);
      \draw[prop] (v5) .. (v7) .. (v4);
      \draw[prop] (v1) .. (v6) .. (v4);
      \draw[prop] (v3) -- (v6);
      \draw[prop] (v5) .. (0.2,0) .. (v6);
      \draw[prop] (v5) .. (-0.2,0) .. (v6);
      \draw[leg] (v2) -- (225:1.2);
      \draw[leg] (v3) -- (270:1.2);
      \fill (v1) circle (1.5pt);
      \fill (v2) circle (1.5pt);
      \fill (v3) circle (1.5pt);
      \fill (v4) circle (1.5pt);
      \fill (v5) circle (1.5pt);
      \fill (v6) circle (1.5pt);
      \fill (v7) circle (1.5pt);
    \end{tikzpicture}
    }}
    =
    \mathcal{K} \left\{  
    \underbrace{\vcenter{\hbox{
    \begin{tikzpicture}[use Hobby shortcut, scale=1.0]
      \draw[prop] (0,0) circle (1.0);
      \coordinate (v1) at (180:1.0);
      \coordinate (v2) at (225:1.0);
      \coordinate (v3) at (270:1.0);
      \coordinate (v4) at (0:1.0);
      \coordinate (v5) at (90:0.5);
      \coordinate (v6) at (270:0.5);
      \coordinate (v7) at (0.6,0.35);
      \coordinate (v7ul) at (-0.6,0.35);
      \coordinate (v7dl) at (-0.6,-0.35);
      \coordinate (v7dr) at (0.6,-0.35);
      \draw[prop] (v1) .. (v7ul) .. (v5);
      \draw[IR] (v5) .. (v7) .. (v4);
      \draw[prop] (v1) .. (v6) .. (v4);
      \draw[prop] (v3) -- (v6);
      \draw[prop] (v5) .. (0.2,0) .. (v6);
      \draw[prop] (v5) .. (-0.2,0) .. (v6);
      \draw[leg] (v2) -- (225:1.2);
      \draw[leg] (v3) -- (270:1.2);
      \fill (v1) circle (1.5pt);
      \fill (v2) circle (1.5pt);
      \fill (v3) circle (1.5pt);
      \fill (v4) circle (1.5pt);
      \fill (v5) circle (1.5pt);
      \fill (v6) circle (1.5pt);
      \fill (v7) circle (1.5pt);
    \end{tikzpicture}       
    }}}_{G}                      
    +
    \left(
    \vcenter{\hbox{
    \begin{tikzpicture}[use Hobby shortcut, scale=1.0]
      \draw[IR] (0,0.45) -- (0,-0.45);
      \fill (0,0) circle (1.5pt);
      \draw (0,0.5) circle (1.5pt);
      \draw (0,-0.5) circle (1.5pt);
    \end{tikzpicture}       
    }}\right)
    \cdot
    \underbrace{\vcenter{\hbox{
    \begin{tikzpicture}[use Hobby shortcut, scale=1.0]
      \draw[prop] (0,0) circle (1.0);
      \coordinate (v1) at (180:1.0);
      \coordinate (v2) at (225:1.0);
      \coordinate (v3) at (270:1.0);
      \coordinate (v4) at (0:1.0);
      \coordinate (v5) at (90:0.5);
      \coordinate (v6) at (270:0.5);
      \coordinate (v7) at (0.6,0.35);
      \coordinate (v7ul) at (-0.6,0.35);
      \coordinate (v7dl) at (-0.6,-0.35);
      \coordinate (v7dr) at (0.6,-0.35);
      \draw[prop] (v1) .. (v7ul) .. (v5);
      \draw[prop] (v1) .. (v6) .. (v4);
      \draw[prop] (v3) -- (v6);
      \draw[prop] (v5) .. (0.2,0) .. (v6);
      \draw[prop] (v5) .. (-0.2,0) .. (v6);
      \draw[leg] (v2) -- (225:1.2);
      \draw[leg] (v3) -- (270:1.2);
      \fill (v1) circle (1.5pt);
      \fill (v2) circle (1.5pt);
      \fill (v3) circle (1.5pt);
      \fill (v4) circle (1.5pt);
      \fill (v5) circle (1.5pt);
      \fill (v6) circle (1.5pt);
    \end{tikzpicture}       
    }}}_{\tilde{G}}    
    \right.\nonumber\\
    %
    %
  & -\mathcal{K}   
    \left( \underbrace{\vcenter{\hbox{
    \begin{tikzpicture}[use Hobby shortcut, scale=1.0]
      \coordinate (v1) at (180:0.5);
      \coordinate (v2) at (0:0.5);
      \draw[prop] (v1) .. (0,0.3) .. (v2);
      \draw[prop] (v1) .. (0,-0.3) .. (v2);
      \draw[leg] (v2) -- (0.7,0);
      \draw[leg] (v1) -- (-0.7,0);
      \fill (v1) circle (1.5pt);
      \fill (v2) circle (1.5pt);
    \end{tikzpicture}       
    }}}_{\gamma_1} \right)\cdot
    \left[
    \underbrace{\vcenter{\hbox{
    \begin{tikzpicture}[use Hobby shortcut, scale=1.0]
      \draw[prop] (0,0) circle (1.0);
      \coordinate (v1) at (180:1.0);
      \coordinate (v2) at (225:1.0);
      \coordinate (v3) at (270:1.0);
      \coordinate (v4) at (0:1.0);
      \coordinate (v5) at (0:0);
      \coordinate (v7) at (0.5,0.35);
      \coordinate (v7ul) at (-0.5,0.35);
      \coordinate (v7dl) at (-0.5,-0.35);
      \coordinate (v7dr) at (0.5,-0.35);
      \draw[prop] (v1) .. (v7ul) .. (v5);
      \draw[prop] (v1) .. (v7dl) .. (v5);
      \draw[IR] (v4) .. (v7) .. (v5);
      \draw[prop] (v4) .. (v7dr) .. (v5);
      \draw[prop] (v3) -- (v5);
      \draw[leg] (v2) -- (225:1.2);
      \draw[leg] (v3) -- (270:1.2);
      \fill (v1) circle (1.5pt);
      \fill (v2) circle (1.5pt);
      \fill (v3) circle (1.5pt);
      \fill (v4) circle (1.5pt);
      \fill (v5) circle (1.5pt);
      \fill (v7) circle (1.5pt);
    \end{tikzpicture}                      
    }}}_{\Gamma_1}
    +
    \left(
    \vcenter{\hbox{
    \begin{tikzpicture}[use Hobby shortcut, scale=1.0]
      \draw[IR] (0,0.45) -- (0,-0.45);
      \fill (0,0) circle (1.5pt);
      \draw (0,0.5) circle (1.5pt);
      \draw (0,-0.5) circle (1.5pt);
    \end{tikzpicture}       
    }}\right)
    \cdot
    \underbrace{\vcenter{\hbox{
    \begin{tikzpicture}[use Hobby shortcut, scale=1.0]
      \draw[prop] (0,0) circle (1.0);
      \coordinate (v1) at (180:1.0);
      \coordinate (v2) at (225:1.0);
      \coordinate (v3) at (270:1.0);
      \coordinate (v4) at (0:1.0);
      \coordinate (v5) at (0:0);
      \coordinate (v7) at (0.5,0.35);
      \coordinate (v7ul) at (-0.5,0.35);
      \coordinate (v7dl) at (-0.5,-0.35);
      \coordinate (v7dr) at (0.5,-0.35);
      \draw[prop] (v1) .. (v7ul) .. (v5);
      \draw[prop] (v1) .. (v7dl) .. (v5);
      \draw[prop] (v4) .. (v7dr) .. (v5);
      \draw[prop] (v3) -- (v5);
      \draw[leg] (v2) -- (225:1.2);
      \draw[leg] (v3) -- (270:1.2);
      \fill (v1) circle (1.5pt);
      \fill (v2) circle (1.5pt);
      \fill (v3) circle (1.5pt);
      \fill (v4) circle (1.5pt);
      \fill (v5) circle (1.5pt);
      \useasboundingbox (-1,-0.5) rectangle (1,0.5);
    \end{tikzpicture}       
    }}}_{\tilde{\Gamma}_1}
    \right]
    \nonumber\\
    %
    %
  & -\KRP   
    \left(
    \underbrace{\vcenter{\hbox{
    \begin{tikzpicture}[use Hobby shortcut, scale=1.0]
      \coordinate (v1) at (180:0.5);
      \coordinate (v2) at (0:0.5);
      \coordinate (v3) at (0,0.3);
      \draw[prop] (v1) .. (0,0.3) .. (v2);
      \draw[prop] (v1) .. (0,-0.3) .. (v2);
      \draw[prop] (v1) .. (-0.1,0.1) .. (v3);
      \draw[leg] (v2) -- (0.7,0);
      \draw[leg] (v1) -- (-0.7,0);
      \fill (v1) circle (1.5pt);
      \fill (v2) circle (1.5pt);
      \fill (v3) circle (1.5pt);
    \end{tikzpicture}       
    }}}_{\gamma_2}
    \right)
    \cdot
    \left[
    \underbrace{\vcenter{\hbox{
    \begin{tikzpicture}[use Hobby shortcut, scale=1.0]
      \draw[prop] (0,0) circle (1.0);
      \coordinate (v1) at (135:1.0);
      \coordinate (v2) at (225:1.0);
      \coordinate (v3) at (270:1.0);
      \coordinate (v4) at (45:1.0);
      \coordinate (v7) at (0,0.7);
      \draw[prop] (v1) .. (0,0.45) .. (v4);
      \draw[IR] (v1) -- (v4);
      \draw[prop] (v3) -- (v1);
      \draw[leg] (v2) -- (225:1.2);
      \draw[leg] (v3) -- (270:1.2);
      \fill (v1) circle (1.5pt);
      \fill (v2) circle (1.5pt);
      \fill (v3) circle (1.5pt);
      \fill (v4) circle (1.5pt);
      \fill (v7) circle (1.5pt);
    \end{tikzpicture}       
    }}}_{\Gamma_2}                      
    +
    \left(
    \vcenter{\hbox{
    \begin{tikzpicture}[use Hobby shortcut, scale=1.0]
      \draw[IR] (0,0.45) -- (0,-0.45);
      \fill (0,0) circle (1.5pt);
      \draw (0,0.5) circle (1.5pt);
      \draw (0,-0.5) circle (1.5pt);
    \end{tikzpicture}       
    }}\right)
    \cdot
    \underbrace{\vcenter{\hbox{
    \begin{tikzpicture}[use Hobby shortcut, scale=1.0]
      \draw[prop] (0,0) circle (1.0);
      \coordinate (v1) at (135:1.0);
      \coordinate (v2) at (225:1.0);
      \coordinate (v3) at (270:1.0);
      \coordinate (v4) at (45:1.0);
      \coordinate (v7) at (0,0.7);
      \draw[prop] (v1) .. (0,0.45) .. (v4);
      \draw[prop] (v3) -- (v1);
      \draw[leg] (v2) -- (225:1.2);
      \draw[leg] (v3) -- (270:1.2);
      \fill (v1) circle (1.5pt);
      \fill (v2) circle (1.5pt);
      \fill (v3) circle (1.5pt);
      \fill (v4) circle (1.5pt);
    \end{tikzpicture}       
    }}}_{\tilde{\Gamma}_2}
    \right]
    \nonumber\\
    %
    %
  & -\KRP   
    \left(
    \underbrace{\vcenter{\hbox{
    \begin{tikzpicture}[use Hobby shortcut, scale=1.0]
      \coordinate (v1) at (180:0.5);
      \coordinate (v2) at (0:0.5);
      \coordinate (v3) at (0,0.3);
      \coordinate (v4) at (0,-0.3);
      \draw[prop] (v1) .. (0,0.3) .. (v2);
      \draw[prop] (v1) .. (0,-0.3) .. (v2);
      \draw[prop] (v1) .. (-0.1,0.1) .. (v3);
      \draw[prop] (v3) -- (v4);
      \draw[leg] (v2) -- (0.7,0);
      \draw[leg] (v1) -- (-0.7,0);
      \fill (v1) circle (1.5pt);
      \fill (v2) circle (1.5pt);
      \fill (v3) circle (1.5pt);
      \fill (v4) circle (1.5pt);
    \end{tikzpicture}       
    }}}_{\gamma_3}
    \right)
    \cdot
    \left[
    \underbrace{\vcenter{\hbox{
    \begin{tikzpicture}[use Hobby shortcut, scale=1.0]
      \coordinate (v1) at (180:0.5);
      \coordinate (v2) at (0:0.5);
      \coordinate (v3) at (0,0.3);
      \draw[prop] (v1) .. (0,0.3) .. (v2);
      \draw[prop] (v1) .. (0,-0.3) .. (v2);
      \draw[prop] (v1) .. (-0.1,0.1) .. (v3);
      \draw[IR] (v3) .. (-0.1,0.5) .. (0,0.6);
      \draw[IR] (v3) .. (0.1,0.5) .. (0,0.6);
      \draw[leg] (v2) -- (0.7,0);
      \draw[leg] (v1) -- (-0.7,0);
      \fill (v1) circle (1.5pt);
      \fill (v2) circle (1.5pt);
      \fill (v3) circle (1.5pt);
      \fill (0,0.6) circle (1.5pt);
    \end{tikzpicture}       
    }}}_{\Gamma_3=0}                      
    +
    \left(
    \vcenter{\hbox{
    \begin{tikzpicture}[use Hobby shortcut, scale=1.0]
      \draw[IR] (0,0.45) -- (0,-0.45);
      \fill (0,0) circle (1.5pt);
      \draw (0,0.5) circle (1.5pt);
      \draw (0,-0.5) circle (1.5pt);
    \end{tikzpicture}       
    }}\right)
    \cdot
    \underbrace{\vcenter{\hbox{
    \begin{tikzpicture}[use Hobby shortcut, scale=1.0]
      \coordinate (v1) at (180:0.5);
      \coordinate (v2) at (0:0.5);
      \coordinate (v3) at (0,0.3);
      \draw[prop] (v1) .. (0,0.3) .. (v2);
      \draw[prop] (v1) .. (0,-0.3) .. (v2);
      \draw[prop] (v1) .. (-0.1,0.1) .. (v3);
      \draw[leg] (v2) -- (0.7,0);
      \draw[leg] (v1) -- (-0.7,0);
      \fill (v1) circle (1.5pt);
      \fill (v2) circle (1.5pt);
      \fill (v3) circle (1.5pt);
    \end{tikzpicture}       
    }}}_{\tilde{\Gamma}_3}
    \right]
    %
    %
  -\KRP   
  \left(
  \underbrace{\vcenter{\hbox{
  \begin{tikzpicture}[use Hobby shortcut, scale=1.0]
    \coordinate (v1) at (180:0.7);
    \coordinate (v4) at (0:0.7);
    \coordinate (v5) at (0,0.3);
    \coordinate (v6) at (0,-0.3);
    \coordinate (v7) at (0.36,0.24);
    \draw[prop] (v1) .. (v5) .. (v4);
    \draw[prop] (v1) .. (v6) .. (v4);
    \draw[prop] (v5) .. (-0.2,0) .. (v6);
    \draw[prop] (v5) .. (0.2,0) .. (v6);
    \draw[prop] (v1) .. (0,-0.5) .. (v4);
    \draw[leg] (v1) -- (-0.9,0);
    \draw[leg] (v7) -- (45:0.7);
    \fill (v1) circle (1.5pt);
    \fill (v4) circle (1.5pt);
    \fill (v5) circle (1.5pt);
    \fill (v6) circle (1.5pt);
    \fill (v7) circle (1.5pt);
  \end{tikzpicture}       
  }}}_{\gamma_4}
  \right)
  \cdot
  \left[
  \underbrace{\vcenter{\hbox{
  \begin{tikzpicture}[use Hobby shortcut, scale=1.0]
    \coordinate (v1) at (180:0.5);
    \coordinate (v2) at (0:0.5);
    \coordinate (v3) at (0,0.3);
    \draw[prop] (v1) .. (0,0.3) .. (v2);
    \draw[prop] (v1) .. (0,-0.3) .. (v2);
    \draw[prop] (v1) .. (-0.1,0.1) .. (v3);
    \draw[leg] (v2) -- (0.7,0);
    \draw[leg] (v1) -- (-0.7,0);
    \fill (v1) circle (1.5pt);
    \fill (v2) circle (1.5pt);
    \fill (v3) circle (1.5pt);
  \end{tikzpicture}       
  }}}_{\Gamma_4}                      
  \right]
  \nonumber\\
  %
  %
  & -\KRP   
    \left(
    \underbrace{\vcenter{\hbox{
    \begin{tikzpicture}[use Hobby shortcut, scale=1.0]
      \coordinate (v1) at (135:1.0);
      \coordinate (v3) at (0:1.0);
      \coordinate (v4) at (45:1.0);
      \coordinate (v5) at (180:1.0);
      \coordinate (v6) at (0,0);
      \draw[prop] (v5) -- (v1);
      \draw[prop] (v1) -- (v4);
      \draw[prop] (v1) -- (v6);
      \draw[prop] (v6) -- (v4);
      \draw[prop] (v3) -- (v4);
      \draw[prop] (v3) -- (v6);
      \draw[prop] (v5) .. (-0.5,0.2) .. (v6);
      \draw[prop] (v5) .. (-0.5,-0.2) .. (v6);
      \draw[leg] (v5) -- (-1.2,0);
      \draw[leg] (v3) -- (1.2,0);
      \fill (v1) circle (1.5pt);
      \fill (v3) circle (1.5pt);
      \fill (v4) circle (1.5pt);
      \fill (v5) circle (1.5pt);
      \fill (v6) circle (1.5pt);
    \end{tikzpicture}       
    }}}_{\gamma_5}
    \right)
    \cdot
    \left[
    \underbrace{\vcenter{\hbox{
    \begin{tikzpicture}[use Hobby shortcut, scale=1.0]
      \coordinate (v1) at (180:0.5);
      \coordinate (v2) at (0:0.5);
      \draw[prop] (v1) .. (0,0.3) .. (v2);
      \draw[prop] (v1) .. (0,-0.3) .. (v2);
      \draw[IR] (v2) .. (0.4,0.2) .. (0.5,0.4);
      \draw[IR] (v2) .. (0.6,0.2) .. (0.5,0.4);
      \draw[leg] (v2) -- (0.7,0);
      \draw[leg] (v1) -- (-0.7,0);
      \fill (v1) circle (1.5pt);
      \fill (v2) circle (1.5pt);
      \fill (0.5,0.4) circle (1.5pt);
    \end{tikzpicture}       
    }}}_{\Gamma_5=0}                      
    +
    \left(
    \vcenter{\hbox{
    \begin{tikzpicture}[use Hobby shortcut, scale=1.0]
      \draw[IR] (0,0.45) -- (0,-0.45);
      \fill (0,0) circle (1.5pt);
      \draw (0,0.5) circle (1.5pt);
      \draw (0,-0.5) circle (1.5pt);
    \end{tikzpicture}       
    }}\right)
    \cdot
    \underbrace{\vcenter{\hbox{
    \begin{tikzpicture}[use Hobby shortcut, scale=1.0]
      \coordinate (v1) at (180:0.5);
      \coordinate (v2) at (0:0.5);
      \draw[prop] (v1) .. (0,0.3) .. (v2);
      \draw[prop] (v1) .. (0,-0.3) .. (v2);
      \draw[leg] (v2) -- (0.7,0);
      \draw[leg] (v1) -- (-0.7,0);
      \fill (v1) circle (1.5pt);
      \fill (v2) circle (1.5pt);
    \end{tikzpicture}       
    }}}_{\tilde{\Gamma}_5}
    \right]
    \nonumber\\
    %
    %
  &\left. -\KRP   
    \left(
    \underbrace{\vcenter{\hbox{
    \begin{tikzpicture}[use Hobby shortcut, scale=1.0]
      \coordinate (v1) at (180:1.0);
      \coordinate (v2) at (0:1.0);
      \coordinate (v3) at (0,1.0);
      \coordinate (vl) at (-0.6,0.5);
      \coordinate (vr) at (0.6,0.5);
      \coordinate (v4) at (0,0);
      \coordinate (v5) at (-0.55,0.55); 
      \coordinate (v6) at (0.3,0.5); 
      \draw[prop] (v1) .. (0,-0.5) .. (v2);
      \draw[prop] (v1) -- (v2);
      \draw[prop] (v1) .. (vl) .. (v3);
      \draw[prop] (v2) .. (vr) .. (v3);
      \draw[prop] (v3) .. (-0.3,0.5) .. (v4);
      \draw[prop] (v3) .. (0.3,0.5) .. (v4);
      \draw[prop] (v4) .. (0.1,0.3) .. (v6);
      \draw[leg] (v2) -- (1.2,0);
      \draw[leg] (v5) -- (-0.75,0.75);
      \fill (v1) circle (1.5pt);
      \fill (v2) circle (1.5pt);
      \fill (v3) circle (1.5pt);
      \fill (v4) circle (1.5pt);
      \fill (v5) circle (1.5pt);
      \fill (v6) circle (1.5pt);
    \end{tikzpicture}       
    }}}_{\gamma_6}
    \right)
    \cdot
    \left[
    \underbrace{\vcenter{\hbox{
    \begin{tikzpicture}[use Hobby shortcut, scale=1.0]
      \draw[IR] (0,0) .. (-0.2,0.5) .. (0,1);
      \draw[IR] (0,0) .. (0.2,0.5) .. (0,1);
      \draw[leg] (0,0) -- (0.2,-0.2);
      \draw[leg] (0,0) -- (-0.2,-0.2);
      \fill (0,0) circle (1.5pt);
      \fill (0,1) circle (1.5pt);
    \end{tikzpicture}       
    }}}_{\Gamma_6=0}                      
    +
    \left(
    \vcenter{\hbox{
    \begin{tikzpicture}[use Hobby shortcut, scale=1.0]
      \draw[IR] (0,0.45) -- (0,-0.45);
      \fill (0,0) circle (1.5pt);
      \draw (0,0.5) circle (1.5pt);
      \draw (0,-0.5) circle (1.5pt);
    \end{tikzpicture}       
    }}\right)
    \right]
    %
    %
    -\KRP   
    \left(
    \underbrace{\vcenter{\hbox{
    \begin{tikzpicture}[use Hobby shortcut, scale=1.0]
      \coordinate (v1) at (180:1.0);
      \coordinate (v2) at (0:1.0);
      \coordinate (v3) at (0,-1.0);
      \coordinate (v4) at (135:1.0);
      \coordinate (v5) at (45:1.0); 
      \coordinate (v6) at (0,0); 
      \draw[prop] (v1) .. (v4) .. (v5) .. (v2);
      \draw[prop] (v1) .. (v3) .. (v2);
      \draw[prop] (v4) -- (v6);
      \draw[prop] (v5) -- (v6);
      \draw[prop] (v3) -- (v6);
      \draw[prop] (v3) -- (v4);
      \draw[prop] (v4) .. (0,0.7) .. (v5);
      \draw[leg] (v1) -- (-1.2,0);
      \draw[leg] (v2) -- (1.2,0);
      \fill (v1) circle (1.5pt);
      \fill (v2) circle (1.5pt);
      \fill (v3) circle (1.5pt);
      \fill (v4) circle (1.5pt);
      \fill (v5) circle (1.5pt);
      \fill (v6) circle (1.5pt);
    \end{tikzpicture}       
    }}}_{\gamma_7}
    \right)
    \cdot
    \underbrace{\vcenter{\hbox{
    \begin{tikzpicture}[use Hobby shortcut, scale=1.0]
      \coordinate (v1) at (180:0.5);
      \coordinate (v2) at (0:0.5);
      \draw[prop] (v1) .. (0,0.3) .. (v2);
      \draw[prop] (v1) .. (0,-0.3) .. (v2);
      \draw[leg] (v2) -- (0.7,0);
      \draw[leg] (v1) -- (-0.7,0);
      \fill (v1) circle (1.5pt);
      \fill (v2) circle (1.5pt);
    \end{tikzpicture}       
    }}}_{\Gamma_7}                      
    \right\}
\end{align}
\bibliography{phi4lc6l}
\end{document}